\documentclass[aps,prl,twocolumn,showpacs,amsmath,amssymb,superscriptaddress,flushbottom,noraggedfooter]{revtex4-1}

\usepackage{graphicx}
\usepackage{subfigure}
\usepackage{amssymb}
\usepackage{bm}

\newcommand{\Rb}{$^{87}$Rb}
\newcommand{\figref}[1]{Fig.~\ref{#1}}

\newcommand{\ket}[1]{\mbox{\ensuremath{\vert #1 \rangle}}}
\newcommand{\xvec}{\boldsymbol{x}}

\newcommand{\kvec}{\boldsymbol{k}}
\newcommand{\lvec}{\boldsymbol{l}}
\newcommand{\Psivec}{\boldsymbol{\Psi}}

\def\imagetop#1{\vtop{\null\hbox{#1}}}

\begin{document}

\title{Quantum noise in three-dimensional BEC interferometry}

\newcommand{\swinburneaffiliation}{Centre for Atom Optics and Ultrafast Spectroscopy, Swinburne University of Technology, Hawthorn, VIC 3122, Australia}
\newcommand{\uqaffiliation}{Physics Department, University of Queensland, Queensland, Australia}

\author{B.~Opanchuk}
\affiliation{Centre for Atom Optics and Ultrafast Spectroscopy, Swinburne University of Technology, Hawthorn, VIC 3122, Australia}
\author{M.~Egorov}
\affiliation{Centre for Atom Optics and Ultrafast Spectroscopy, Swinburne University of Technology, Hawthorn, VIC 3122, Australia}
\author{S.~Hoffmann}
\affiliation{Physics Department, University of Queensland, Queensland, Australia}
\author{A.~Sidorov}
\affiliation{Centre for Atom Optics and Ultrafast Spectroscopy, Swinburne University of Technology, Hawthorn, VIC 3122, Australia}
\author{P.~D.~Drummond}
\affiliation{Centre for Atom Optics and Ultrafast Spectroscopy, Swinburne University of Technology, Hawthorn, VIC 3122, Australia}

\begin{abstract}
We develop a theory of quantum fluctuations and squeezing in a three-dimensional Bose-Einstein condensate atom interferometer with nonlinear losses.
We use stochastic equations in a truncated Wigner representation to treat quantum noise.
Our approach includes the multi-mode spatial evolution of spinor components and describes the many-body dynamics of a mesoscopic quantum system.
\end{abstract}


\date{\today}

\maketitle

Atom interferometry is an important quantum technology
at the heart of many proposed future applications of ultra-cold atomic physics.
Bose-Einstein condensates (BECs) or atom lasers are macroscopic quantum objects and have potential advantages as interferometric detectors and sensors,
provided one can precisely extract atomic phase information.
However, unlike photons, atoms can interact strongly, causing dephasing and loss of interference fringes.
An intimate understanding of quantum many-body dynamics is the key to calculating
interaction-induced dephasing in the measurement process.
This is essential for a quantitative theory of atom interferometry.

In this Letter we present a simple, yet quantitatively accurate theoretical approach
to simulating the dynamics and evaluating limits of atom interferometry at large atom number,
using a truncated Wigner representation~\cite{Drummond1993,Steel1998,Sinatra2002}.
This method extends the conventional Gross-Pitaevskii equations
describing a Bose condensate to include quantum noise effects,
including noise due to linear and nonlinear losses.
The theory allows the accurate inclusion of quantum fluctuations due to nonlinear losses,
which is a dominant effect when atom numbers are increased to improve fringe visibility.

Importantly, we can clearly demonstrate where fringe visibility is driven by quantum fluctuations,
and where it is driven by trap inhomogeneity and dynamical effects,
in order to choose optimal conditions for quantum noise reduction and spin squeezing.
These calculations are a first step towards understanding mesoscopic superpositions and entanglement in ultra-cold atomic gases.
An advantage of our method compared to the variational approaches used elsewhere~\cite{Li2009,Sakmann2009}
is that it allows us to treat a large number of independent field modes and particles,
thus including degrees of freedom that are excited due to
collisional and nonlinear loss dynamics~\cite{Norrie2005,Deuar2007}.
Our theory can be readily extended to include finite temperature initial conditions~\cite{Steel1998,Isella2006},
which will be treated elsewhere.
Nonlinear losses and finite temperature effects can be also described within the confines of the variational approach~\cite{Li2008,Sinatra2011}.

Quantum phase-diffusion is defined as the phase noise induced by number fluctuations
which are conjugate to phase.
This is a fundamental feature of BEC interferometry, and can only be removed when there are no interactions.
However, there are other reasons for decoherence, which are also important.
The approach used here captures all three significant features of atom
interferometry that can result in decoherence: phase-diffusion, losses,
and trap inhomogeneity effects.
The results given in this paper are applicable to simulations where the
atom number per lattice point or mode is large.
We focus on trapped BEC interferometry, for definiteness, although the method has a more general applicability.

We start by assuming that the BEC has $s$-wave interactions,
together with Markovian losses due to $n$-body collisions.
We employ a master equation together with the Wigner-Moyal quantum phase-space representation~\cite{Gardiner2004}
and a truncation of third and higher-order derivatives in the equations of motion.
If we regard the commonly used Gross-Pitaevskii equation as a classical,
first approximation to mean-field condensate dynamics,
the truncated Wigner approach is best thought of as the second term in an expansion in inverse particle number.
This truncation has been shown to be valid in the limit of large particle
number~\cite{Drummond1993,Steel1998,Sinatra2002}.
It has been further tested by comparison with the exact positive-P simulation method.
The truncated Wigner method is particularly useful in low-dimensional and
trap environments, where it has successfully predicted quantum squeezing
and phase-diffusion effects, in good agreement with dynamical experiments
in photonic quantum soliton propagation~\cite{Carter1987,Corney2008}.

In the present Letter, we treat an ultra-cold,
interacting multi-component spinor Bose gas in $D$ effective dimensions.
The basic Hamiltonian is easily expressed using quantum fields
$\widehat{\Psi}_j^{\dagger}(\xvec)$ and $\widehat{\Psi}_j(\xvec)$,
where $\widehat{\Psi}_j^{\dagger}(\xvec)$ creates a bosonic atom of spin $j$
at location $\xvec$, and $\widehat{\Psi}_j(\xvec)$ destroys one;
the commutators are
$[\widehat{\Psi}_j(\xvec),\widehat{\Psi}_k^{\dagger}(\xvec^\prime)] =
\delta^{(D)}(\xvec-\xvec^\prime)\delta_{jk}\,\,.$
The resulting physics of a dilute, low-temperature Bose gas
is well-described in the $s$-wave scattering limit by an effective Hamiltonian
with contact interactions and external potentials:
\begin{equation}
	\hat{H} / \hbar = \int d^{D}\xvec \left\{
		\widehat{\Psi}_j^{\dagger} K_{jk} \widehat{\Psi}_k +
		\frac{U_{jk}}{2} \widehat{\Psi}_j^{\dagger} \widehat{\Psi}_k^{\dagger}
		\widehat{\Psi}_k \widehat{\Psi}_j
	\right\}.
\end{equation}
Here we omit the field argument $(\xvec)$ for brevity,
and use the Einstein summation convention of summing over repeated indices.
$K_{jk}$ is the single-particle Hamiltonian:
\begin{equation}
	K_{jk} = \left( -\frac{\hbar}{2m} \nabla^2 + \omega_j + V_j(\xvec) / \hbar \right) \delta_{jk} +
		\tilde{\Omega}_{jk}(t),
\end{equation}
where $m$ is the atomic mass, $V_j$ is the external trapping potential for spin $j$,
$\omega_j$ is the internal energy of spin $j$,
$\tilde{\Omega}_{jk}$ represents a time-dependent coupling
that is used to rotate one spin projection into another,
and $U_{jk}$ is the atom-atom interaction term.
Thus, $n_j = \langle \widehat{\Psi}_j^{\dagger} \widehat{\Psi}_j \rangle$
is the spin-$j$ atomic density.
For a dilute gas at low enough temperatures,
$U_{jk}=4\pi\hbar a_{jk} / m$, where $a_{jk}$ is the $s$-wave scattering length in three dimensions.
Here we assume a momentum cutoff $k_{c} \ll 1 / a_{jk}$,
otherwise the couplings must be renormalized~\cite{Sinatra2002}.

We proceed by using a stochastic phase-space method that allows a numerical
simulation of the quantum dynamics~\cite{Drummond1993,Steel1998,Hoffmann2008}.
Defining a Wigner function $W(\Psivec)$, where $\Psi$
is a c-number field corresponding to the quantum field $\hat{\Psi}$, this has a unitary time-evolution equation:
\begin{equation}
	\frac{\partial W}{\partial t} = \int d^D\xvec \left\{
		- \frac{\delta}{\delta\Psi_j} A_j
		- \frac{\delta}{\delta\Psi_j^*}A_j^*
		+ \mbox{O} \left[ \frac{\delta^3}{\delta\Psi_j^3} \right]
	\right\} W.
\end{equation}
Next, higher derivative terms of type $\mbox{O} \left[ \delta^3 / \delta\Psi_j^3 \right]$ are truncated.
This approximation neglects higher-order terms in an expansion in $1 / \sqrt{N}$,
and is therefore valid in the limit of $N \gg M$
where $N$ is the atom number and $M$ is the number of low-energy modes included~\cite{Drummond1993,Sinatra2002,Norrie2006}.
In free-space calculations it is important to maintain this mode truncation.
In the relevant limits where the technique is applicable, the equations
simply reduce to Gross-Pitaevskii equations with Gaussian fluctuations
of the initial conditions:
\begin{equation}
\label{eqn:SDE-1}
	\frac{d\Psi_j}{dt} = -i \left(
		K_{jk} \Psi_k + U_{jk} \lvert \Psi_k \rvert^2 \Psi_j
	\right).
\end{equation}
For initial conditions in interferometry it is usually sufficient to consider
a coherent state amplitude $\Psi_s^c$,
corresponding to a typical initial state with Poissonian number fluctuations,
as produced by a beam-splitter.
In this case the initial Wigner amplitude has a Gaussian random distribution, with
$\Psi_j(\xvec, t_0) = \Psi_j^c(\xvec) + \Delta \Psi_j(\xvec)$, where:
$\left\langle \Delta \Psi_j(\xvec) \Delta \Psi_k^*(\xvec^{\prime}) \right\rangle =
\delta_{jk} \delta^D(\xvec - \xvec^{\prime}) / 2.$
This initial noise is necessary because the Wigner representation generates
symmetrically ordered correlation functions, and includes vacuum fluctuations.
For greater accuracy, the initial state can be modified to account for
initial  correlations, thermal noise, or additional fluctuations.
If normal ordered correlations are measured, one has to express them
as a sum of symmetrically ordered terms.

This includes all the known nonlinear quantum noise effects of quantum dynamics,
like phase diffusion, entanglement and quantum squeezing, in the limit
of large particle number.
The initial noise terms do not occur in the semi-classical Gross-Pitaevskii
approximation, which is therefore unable to predict these effects.
Thus, while the lossless equations are identical to the Gross-Pitaevskii
equations, the inclusion of initial noise terms together with nonlinear
interactions leads to quantum phase-diffusion.
Such methods can be used for either free-space or trapped atom interferometry,
provided there is an appropriate mode truncation.

Additional quantum noise enters from the effects of damping and losses,
due to the fluctuation-dissipation theorem.
These effects are important at high densities in atomic traps.
They can be included via an additional Markovian master equation~\cite{Jack2002}
defined so that,
\begin{equation}
	\frac{d\hat{\rho}}{dt} =
		- \frac{i}{\hbar} \left[ \hat{H}, \hat{\rho} \right]
		+ \sum_{n,\lvec} \kappa_{\lvec}^{(n)} \int d^{D}\xvec
			\mathcal{L}_{\lvec}^{(n)} \left[ \hat{\rho} \right],
\end{equation}
where $n$ is the number of interacting particles,
$\lvec = (l_1, l_2, \ldots, l_n)$ is a vector indicating the spins that are coupled,
and we have introduced local Liouville loss terms,
\begin{equation}
	\mathcal{L}_{\lvec}^{(n)} \left[ \hat{\rho} \right] =
		2\hat{O}_{\lvec}^{(n)} \hat{\rho} \hat{O}_{\lvec}^{(n)\dagger}
		- \hat{O}_{\lvec}^{(n)\dagger} \hat{O}_{\lvec}^{(n)} \hat{\rho}
		- \hat{\rho} \hat{O}_{\lvec}^{(n)\dagger} \hat{O}_{\lvec}^{(n)}.
\end{equation}
The reservoir coupling operators $\hat{O}_{\lvec}^{(n)}$ are the distinct $n$-fold products of local field annihilation operators,
$\hat{O}_{\lvec}^{(n)} = \hat{O}_{\lvec}^{(n)} (\widehat{\Psivec}) =
	\widehat{\Psi}_{l_{1}} (\xvec)
	\widehat{\Psi}_{l_{2}} (\xvec) \ldots
	\widehat{\Psi}_{l_{n}} (\xvec),$
describing local $n$-body collision losses.

After transforming these new terms to evolution equations for the Wigner distribution, the drift term $A_j$
changes the Gross-Pitaevskii evolution to include nonlinear damping, while
the next terms in the evolution equation give rise to additional Fokker-Planck
diffusion terms associated with quantum noise from the loss reservoirs,
given by:
\begin{equation}
	\frac{\delta^{2}}{\delta\Psi_j\delta\Psi_k^{*}} \left\{
		\sum_{n,\lvec} \kappa_{\lvec}^{(n)}
			\frac{\partial O_{\lvec}^{(n)*}}{\partial\Psi_j^{*}}
			\frac{\partial O_{\lvec}^{(n)}}{\partial\Psi_k}
		\right\} W.
\end{equation}

This leads to a stochastic equation:
\begin{equation}
\label{eqn:SDE}
	\frac{d\Psi_j}{dt} =
		- i\left( K_{jk} \Psi_k + U_{jk} \lvert \Psi_k \rvert^{2} \Psi_j \right)
		- \Gamma_j
		+ \sum_{n,\lvec} \beta_{\lvec,j}^{(n)} \zeta_{\lvec}^{(n)}(\xvec,t),
\end{equation}
where the nonlinear loss has the form:
\begin{equation}
	\Gamma_j = \sum_{n,\lvec}
		\kappa_{\lvec}^{(n)}
		\frac{\partial O_{\lvec}^{(n)*} (\Psivec)}{\partial\Psi_j^{*} (\xvec)}
		O_{\lvec}^{(n)}(\xvec),
\end{equation}
and $\zeta_{\lvec}^{(n)}(\xvec, t)$ is a corresponding complex,
stochastic delta-correlated Gaussian noise with
\begin{equation}
	\left\langle
		\zeta_{\lvec}^{(n)} (\xvec,t) \zeta_{\kvec}^{(m)*}(\xvec^\prime, t^\prime)
	\right\rangle =
	\delta_{\lvec \kvec} \delta^{nm} \delta^{D} \left(
		\xvec - \xvec^\prime
	\right)
	\delta \left( t - t^\prime \right).
\end{equation}
The multiplicative noise coefficient
\begin{equation}
	\beta_{\lvec,j}^{(n)} \left( \Psivec \right) =
	\sqrt{\kappa_{\lvec}^{(n)}}
	\frac{\partial O_{\lvec}^{(n)}}{\partial\Psi_j}
\end{equation}
is a fluctuation-dissipation term,
so that the Wigner variables remain equivalent to the corresponding operators.

The loss coefficients in equations~(\ref{eqn:SDE}) can be converted to the conventional form,
which is defined using atom number losses:
\begin{equation}
	\dot{n_j} = - \gamma^{(n)}_{\lvec,j} n^{m_1}_1 n^{m_2}_2 \ldots ,
\end{equation}
where $n_j$ is the density of component $j$ and $m_j$
is the number of spin-$j$ atoms lost in the collision.
The conversion can be carried out as $\gamma^{(n)}_{\lvec,j} = 2 m_j \kappa^{(n)}_{\lvec}$.

In this work we use a basis of plane waves in the volume $V$,
and the density of component $j$ is calculated as a probabilistic average:
\begin{equation}
\label{eqn:wigner-density}
	n_j (\xvec)
		= \langle \Psi^*_j (\xvec) \Psi_j (\xvec) \rangle_{\mathrm{paths}} - \frac{M}{2V}.
\end{equation}
Here we use the fact that the approximate Wigner function is a probability distribution
equivalent to an averaged sum over different simulation paths.

\begin{figure}
	\begin{tabular}{l l}
	\imagetop{\hspace*{0.44in}\includegraphics[width=2.6in]{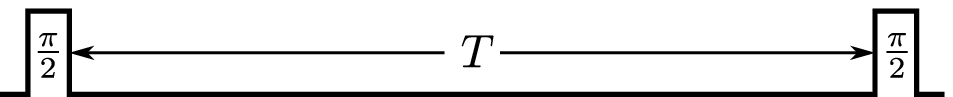}} & \imagetop{(a)} \\
	\imagetop{\includegraphics{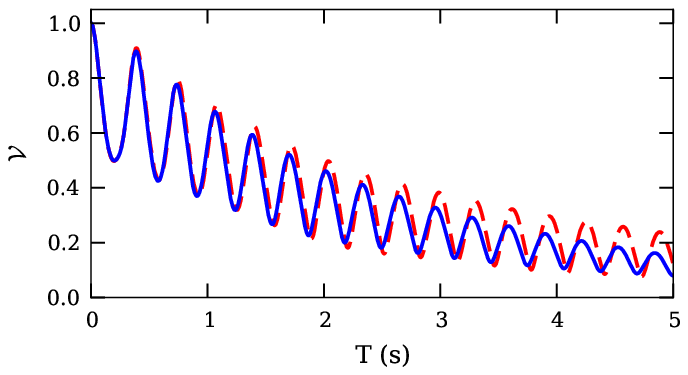}} & \imagetop{(b)} \\
	\imagetop{\hspace*{0.44in}\includegraphics[width=2.6in]{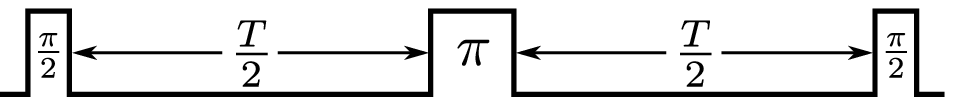}} & \imagetop{(c)} \\
	\imagetop{\includegraphics{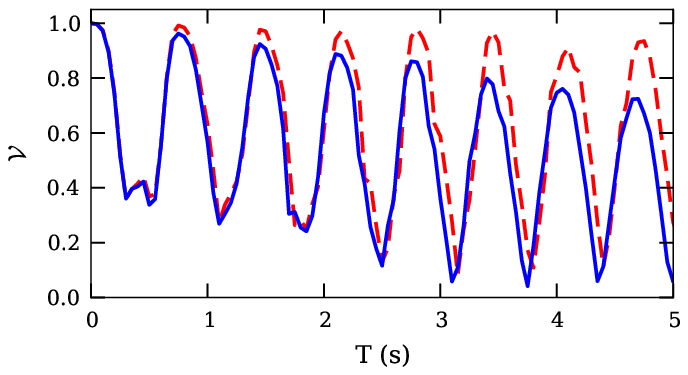}} & \imagetop{(d)}
	\end{tabular}

	\caption{(color online) Timeline of the experiment for Ramsey (a) and Ramsey with spin echo (c),
	(b) and (d) are the simulated plots of interferometric visibility.
	Classical GPE (red dashed lines) and Wigner calculations (blue solid lines)
	are shown. $N = 5.5 \times 10^4$,
	$\omega_x = \omega_y = 2 \pi \times 97.0\,\mathrm{Hz}$,
	$\omega_z = 2 \pi \times 11.69\,\mathrm{Hz}$,
	$a_{11} = 100.4\,a_0$, $a_{12} = 97.993\,a_0$, $a_{22} = 95.57\,a_0$~\cite{Egorov2011},
	$a_0$ is the Bohr radius.
	Nonlinear atomic losses:
	$\gamma^{(3)}_{111} = 5.4 \times 10^{-30}\,\mathrm{cm^6/s}$~\cite{Mertes2007},
	$\gamma^{(2)}_{12} = 1.51 \times 10^{-14}\,\mathrm{cm^3/s}$,
	$\gamma^{(2)}_{22} = 8.1 \times 10^{-14}\,\mathrm{cm^3/s}$~\cite{Egorov2011}.}

	\label{fig:visibility}
\end{figure}

To illustrate the applications of this method we consider recent interferometry
experiments with a two-component BEC involving two hyperfine states
${\ket{F=1,\, m_F=-1}}$ and ${\ket{F=2,\, m_F=+1}}$ in \Rb~\cite{Egorov2011}.
A conventional Ramsey sequence (\figref{fig:visibility},~(a)) has been used
with a BEC confined in a cigar-shaped magnetic trap with the frequencies $(97.0, 97.0, 11.69)\,\mathrm{Hz}$
in a bias magnetic field of $3.23\,\mathrm{G}$, so that magnetic field dephasing is largely eliminated~\cite{Hall1998}.
The first $\pi/2$ pulse prepares a non-equilibrium superposition of states ${\ket{1,-1}}$ and ${\ket{2,+1}}$
and the spatial modes of two components periodically separate and merge again~\cite{Mertes2007}.
The spatially-separated spin components evolve differently, as they have
different scattering lengths.
As a result, these collective oscillations lead to periodic dephasing and
self-rephasing of the BEC components, clearly visible in both GPE and Wigner
simulations of interference fringe visibility
$\mathcal{V}$ (\figref{fig:visibility},~(b)).
Asymmetric losses of two states are one cause of the contrast decay.
This can be partially compensated by the application of a spin echo pulse
mid-way through the evolution (\figref{fig:visibility},~(c)).
The GPE simulations wrongly predict (dashed lines) that visibility is largely
recovered at long evolution times using the spin echo method.
However, the addition of quantum noise (solid line) via the Wigner simulations
noticeably speeds up the visibility decay even with a spin echo pulse present.
This is in agreement with experimental observations, and shows that these
effects play a significant part in the decay of visibility, even for
large particle numbers.

The important feature of these quantum dynamical simulations
is that they are able to treat large numbers of atoms (55,000 in this case),
while correctly tracking all the quantum noise sources, and also extending the simulations to long time-scales.
Both of these features, large atom numbers and long time-scales,
are essential ingredients to accurate interferometric measurements.
The simulations give accurate predictions despite large, multi-mode dynamical motion in three dimensions
and substantial losses of most of the condensate atoms~\cite{Egorov2011}.
On longer time-scales, the experimental accuracy is limited by technical noises, and we have no data for comparisons.

\begin{figure}
	\includegraphics{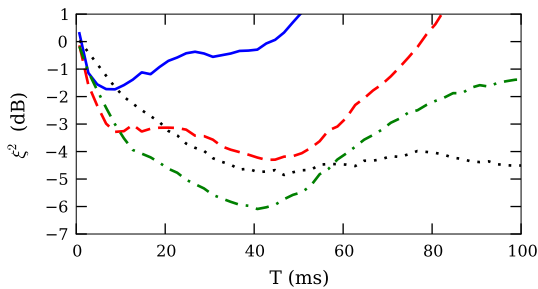}

	\caption[]{(color online)
	Wigner simulations of squeezing in the vicinity of $9.1\,\mathrm{G}$ Feshbach resonance in \Rb.
	Simulation parameters are the same as in~\figref{fig:visibility},
	except for the inter-component scattering length and corresponding loss rate~\cite{Kaufman2009}:
	$a_{12} = 80.0\,a_0$, $\gamma^{(2)}_{12} = 3.88 \times 10^{-12}\,\mathrm{cm^3/s}$ (blue solid line),
	$a_{12} = 85.0\,a_0$, $\gamma^{(2)}_{12} = 1.95 \times 10^{-12}\,\mathrm{cm^3/s}$ (red dashed line),
	$a_{12} = 90.0\,a_0$, $\gamma^{(2)}_{12} = 7.13 \times 10^{-13}\,\mathrm{cm^3/s}$ (green dash-dotted line) and
	$a_{12} = 95.0\,a_0$, $\gamma^{(2)}_{12} = 8.54 \times 10^{-14}\,\mathrm{cm^3/s}$ (black dotted line)}

	\label{fig:squeezing}
\end{figure}

The Wigner method is able to predict not only the average values of different observables,
but their variances too.
As an example, we now calculate the degree of squeezing $\xi^2$, introduced in~\cite{Wineland1994,Sorensen2001}.
This allows the analysis of experiments beyond the usual shot noise limit~\cite{Riedel2010,Gross2010}.
If we define spin component operators as
$\hat{S}_{x} = [ \hat{S}_+ + \hat{S}_+^\dagger ] / 2 $,
$\hat{S}_{y} = i [ \hat{S}_+ - \hat{S}_+^\dagger ] / 2 $ and
\begin{equation}
\begin{split}
	\hat{S}_+ & = \int d^D\xvec \left[
		\widehat{\Psi}^\dagger_2 (\xvec) \widehat{\Psi}_1 (\xvec)
	\right], \\
	\hat{S}_z & = \frac{1}{2} \int d^D\xvec \left[
		\widehat{\Psi}^\dagger_1 (\xvec) \widehat{\Psi}_1 (\xvec)
		- \widehat{\Psi}^\dagger_2 (\xvec) \widehat{\Psi}_2 (\xvec)
	\right],
\end{split}
\end{equation}
the average values of spin vector components can be calculated similarly to the average density in
equation~(\ref{eqn:wigner-density}).
Now we can shift the centre of coordinates to the end of the vector
$\langle \boldsymbol{S} \rangle = \{ \langle \hat{S}_x \rangle, \langle \hat{S}_y \rangle, \langle \hat{S}_z \rangle \}$
and rotate it, making the $z^\prime$ axis parallel to $\boldsymbol{S}$ and choosing the remaining axes $x^\prime$ and $y^\prime$ so that
$\Delta \hat{S}^2_{y^\prime} $ is minimized.

Second-order moments of spin operators are represented using fourth-order moments of field operators
and, therefore, can be calculated using wavefunctions in Wigner representation.
In this new coordinate system the squeezing parameter is expressed simply as
\begin{equation}
\label{eqn:squeezing}
	\xi^2 = \frac{N \Delta \hat{S}^2_{y^\prime}}{\langle S \rangle^2},
\end{equation}
where $N$ is the total number of atoms.
The formula for the squeezing parameter in the original coordinate system can be found elsewhere~\cite{Li2009}.
The squeezing parameter serves as a measure of atomic entanglement in the condensate,
when the atoms are entangled if $\xi^2 < 1$~\cite{Sorensen2001}.

As an illustration of the power of the Wigner method and of the effect of two-body losses we consider the temporal evolution of the squeezing parameter in a Ramsey interferometer for optically trapped two-component BEC in \Rb\ (states ${\ket{F=1,\, m_F=+1}}$ and ${\ket{F=2,\, m_F=-1}}$) near a Feshbach resonance at $9.1\,\mathrm{G}$ (\figref{fig:squeezing}).
Different inter-component scattering lengths (corresponding to different magnetic fields)
were used in order to find the optimal regime with significant squeezing.
The value $a_{12} = 80 a_0$ provides stronger nonlinear interactions, but larger two-body losses quickly eliminate the squeezing effect.
Feshbach tuning to $a_{12} = 90 a_0$ ensures the best squeezing ($-6\,\mathrm{dB}$ at $40\,\mathrm{ms}$), whereas long lasting squeezing is predicted for $a_{12} = 95 a_0$.

These simulations predict the degree of quantum noise-reduction, which
is a technique that can be used to improve precision quantum interferometry.
The quantum noise is initially reduced due to a stretching and rotation
of the quantum noise ellipse, similar to that found in quantum soliton
squeezing~\cite{Carter1987,Drummond1993a}.
For lossless environments, this effect is optimized when the cross-species
scattering length is maximally different to the intra-species scattering
length, which is found near the Feshbach resonance.
However, there are competing nonlinear loss effects at the Feshbach resonance,
which means that some detuning of the magnetic field is essential to reduce
these detrimental losses.
Subsequently, the squeezing is destroyed in time as the two fields recombine
and interfere with each other.
Importantly, these quantum squeezing calculations indicate conditions that
will allow this macroscopic quantum effect to be experimentally
observed in ultra-cold atomic BEC for much larger atom
numbers than calculated previously.

In conclusion, the truncated Wigner-Moyal quantum phase-space theory is
well-suited for a variety of calculations of many-body dynamics of quantum degenerate gases.
Although the Wigner function is not a true probability, it becomes one
to a good approximation in the limit of large particle number, so that
a probabilistic (random) sampling of phase-space trajectories can be used
to treat quantum fluctuations.
Furthermore, it is straightforward to treat a spatially inhomogeneous,
many-mode problem as found in an extended 3-dimensional environment.
Initial conditions such as a coherent beam-splitter as employed in typical
atom interferometry experiments can be readily included,
with proper accounting for both quantum effects and nonlinear losses in the condensate.

\bibliography{qsim.bib}

\begin{thebibliography}{25}%
\makeatletter
\providecommand \@ifxundefined [1]{%
 \@ifx{#1\undefined}
}%
\providecommand \@ifnum [1]{%
 \ifnum #1\expandafter \@firstoftwo
 \else \expandafter \@secondoftwo
 \fi
}%
\providecommand \@ifx [1]{%
 \ifx #1\expandafter \@firstoftwo
 \else \expandafter \@secondoftwo
 \fi
}%
\providecommand \natexlab [1]{#1}%
\providecommand \enquote  [1]{``#1''}%
\providecommand \bibnamefont  [1]{#1}%
\providecommand \bibfnamefont [1]{#1}%
\providecommand \citenamefont [1]{#1}%
\providecommand \href@noop [0]{\@secondoftwo}%
\providecommand \href [0]{\begingroup \@sanitize@url \@href}%
\providecommand \@href[1]{\@@startlink{#1}\@@href}%
\providecommand \@@href[1]{\endgroup#1\@@endlink}%
\providecommand \@sanitize@url [0]{\catcode `\\12\catcode `\$12\catcode
  `\&12\catcode `\#12\catcode `\^12\catcode `\_12\catcode `\%12\relax}%
\providecommand \@@startlink[1]{}%
\providecommand \@@endlink[0]{}%
\providecommand \url  [0]{\begingroup\@sanitize@url \@url }%
\providecommand \@url [1]{\endgroup\@href {#1}{\urlprefix }}%
\providecommand \urlprefix  [0]{URL }%
\providecommand \Eprint [0]{\href }%
\providecommand \doibase [0]{http://dx.doi.org/}%
\providecommand \selectlanguage [0]{\@gobble}%
\providecommand \bibinfo  [0]{\@secondoftwo}%
\providecommand \bibfield  [0]{\@secondoftwo}%
\providecommand \translation [1]{[#1]}%
\providecommand \BibitemOpen [0]{}%
\providecommand \bibitemStop [0]{}%
\providecommand \bibitemNoStop [0]{.\EOS\space}%
\providecommand \EOS [0]{\spacefactor3000\relax}%
\providecommand \BibitemShut  [1]{\csname bibitem#1\endcsname}%
\let\auto@bib@innerbib\@empty
\bibitem [{\citenamefont {Drummond}\ and\ \citenamefont
  {Hardman}(1993)}]{Drummond1993}%
  \BibitemOpen
  \bibfield  {author} {\bibinfo {author} {\bibfnamefont {P.~D.}\ \bibnamefont
  {Drummond}}\ and\ \bibinfo {author} {\bibfnamefont {A.~D.}\ \bibnamefont
  {Hardman}},\ }\href {\doibase 10.1209/0295-5075/21/3/005} {\bibfield
  {journal} {\bibinfo  {journal} {Europhys. Lett.}\ }\textbf {\bibinfo {volume}
  {21}},\ \bibinfo {pages} {279} (\bibinfo {year} {1993})}\BibitemShut
  {NoStop}%
\bibitem [{\citenamefont {Steel}\ \emph {et~al.}(1998)\citenamefont {Steel}
  \emph {et~al.}}]{Steel1998}%
  \BibitemOpen
  \bibfield  {author} {\bibinfo {author} {\bibfnamefont {M.}~\bibnamefont
  {Steel}} \emph {et~al.},\ }\href {\doibase 10.1103/PhysRevA.58.4824}
  {\bibfield  {journal} {\bibinfo  {journal} {Phys. Rev. A}\ }\textbf {\bibinfo
  {volume} {58}},\ \bibinfo {pages} {4824} (\bibinfo {year}
  {1998})}\BibitemShut {NoStop}%
\bibitem [{\citenamefont {Sinatra}\ \emph {et~al.}(2002)\citenamefont
  {Sinatra}, \citenamefont {Lobo},\ and\ \citenamefont {Castin}}]{Sinatra2002}%
  \BibitemOpen
  \bibfield  {author} {\bibinfo {author} {\bibfnamefont {A.}~\bibnamefont
  {Sinatra}}, \bibinfo {author} {\bibfnamefont {C.}~\bibnamefont {Lobo}}, \
  and\ \bibinfo {author} {\bibfnamefont {Y.}~\bibnamefont {Castin}},\ }\href
  {\doibase 10.1088/0953-4075/35/17/301} {\bibfield  {journal} {\bibinfo
  {journal} {J. Phys. B}\ }\textbf {\bibinfo {volume} {35}},\ \bibinfo {pages}
  {3599} (\bibinfo {year} {2002})}\BibitemShut {NoStop}%
\bibitem [{\citenamefont {Li}\ \emph {et~al.}(2009)\citenamefont {Li} \emph
  {et~al.}}]{Li2009}%
  \BibitemOpen
  \bibfield  {author} {\bibinfo {author} {\bibfnamefont {Y.}~\bibnamefont {Li}}
  \emph {et~al.},\ }\href {\doibase 10.1140/epjb/e2008-00472-6} {\bibfield
  {journal} {\bibinfo  {journal} {Eur. Phys. J. B}\ }\textbf {\bibinfo {volume}
  {68}},\ \bibinfo {pages} {365} (\bibinfo {year} {2009})}\BibitemShut
  {NoStop}%
\bibitem [{\citenamefont {Sakmann}\ \emph {et~al.}(2009)\citenamefont {Sakmann}
  \emph {et~al.}}]{Sakmann2009}%
  \BibitemOpen
  \bibfield  {author} {\bibinfo {author} {\bibfnamefont {K.}~\bibnamefont
  {Sakmann}} \emph {et~al.},\ }\href {\doibase 10.1103/PhysRevLett.103.220601}
  {\bibfield  {journal} {\bibinfo  {journal} {Phys. Rev. Lett.}\ }\textbf
  {\bibinfo {volume} {103}},\ \bibinfo {pages} {220601} (\bibinfo {year}
  {2009})}\BibitemShut {NoStop}%
\bibitem [{\citenamefont {Norrie}\ \emph {et~al.}(2005)\citenamefont {Norrie},
  \citenamefont {Ballagh},\ and\ \citenamefont {Gardiner}}]{Norrie2005}%
  \BibitemOpen
  \bibfield  {author} {\bibinfo {author} {\bibfnamefont {A.}~\bibnamefont
  {Norrie}}, \bibinfo {author} {\bibfnamefont {R.}~\bibnamefont {Ballagh}}, \
  and\ \bibinfo {author} {\bibfnamefont {C.}~\bibnamefont {Gardiner}},\ }\href
  {\doibase 10.1103/PhysRevLett.94.040401} {\bibfield  {journal} {\bibinfo
  {journal} {Phys. Rev. Lett.}\ }\textbf {\bibinfo {volume} {94}},\ \bibinfo
  {pages} {040401} (\bibinfo {year} {2005})}\BibitemShut {NoStop}%
\bibitem [{\citenamefont {Deuar}\ and\ \citenamefont
  {Drummond}(2007)}]{Deuar2007}%
  \BibitemOpen
  \bibfield  {author} {\bibinfo {author} {\bibfnamefont {P.}~\bibnamefont
  {Deuar}}\ and\ \bibinfo {author} {\bibfnamefont {P.}~\bibnamefont
  {Drummond}},\ }\href {\doibase 10.1103/PhysRevLett.98.120402} {\bibfield
  {journal} {\bibinfo  {journal} {Phys. Rev. Lett.}\ }\textbf {\bibinfo
  {volume} {98}},\ \bibinfo {pages} {120402} (\bibinfo {year}
  {2007})}\BibitemShut {NoStop}%
\bibitem [{\citenamefont {Isella}\ and\ \citenamefont
  {Ruostekoski}(2006)}]{Isella2006}%
  \BibitemOpen
  \bibfield  {author} {\bibinfo {author} {\bibfnamefont {L.}~\bibnamefont
  {Isella}}\ and\ \bibinfo {author} {\bibfnamefont {J.}~\bibnamefont
  {Ruostekoski}},\ }\href {\doibase 10.1103/PhysRevA.74.063625} {\bibfield
  {journal} {\bibinfo  {journal} {Phys. Rev. A}\ }\textbf {\bibinfo {volume}
  {74}},\ \bibinfo {pages} {063625} (\bibinfo {year} {2006})}\BibitemShut
  {NoStop}%
\bibitem [{\citenamefont {Li}\ \emph {et~al.}(2008)\citenamefont {Li},
  \citenamefont {Castin},\ and\ \citenamefont {Sinatra}}]{Li2008}%
  \BibitemOpen
  \bibfield  {author} {\bibinfo {author} {\bibfnamefont {Y.}~\bibnamefont
  {Li}}, \bibinfo {author} {\bibfnamefont {Y.}~\bibnamefont {Castin}}, \ and\
  \bibinfo {author} {\bibfnamefont {A.}~\bibnamefont {Sinatra}},\ }\href
  {\doibase 10.1103/PhysRevLett.100.210401} {\bibfield  {journal} {\bibinfo
  {journal} {Phys. Rev. Lett.}\ }\textbf {\bibinfo {volume} {100}},\ \bibinfo
  {pages} {210401} (\bibinfo {year} {2008})}\BibitemShut {NoStop}%
\bibitem [{\citenamefont {Sinatra}\ \emph {et~al.}(2011)\citenamefont {Sinatra}
  \emph {et~al.}}]{Sinatra2011}%
  \BibitemOpen
  \bibfield  {author} {\bibinfo {author} {\bibfnamefont {A.}~\bibnamefont
  {Sinatra}} \emph {et~al.},\ }\href {\doibase 10.1103/PhysRevLett.107.060404}
  {\bibfield  {journal} {\bibinfo  {journal} {Phys. Rev. Lett.}\ }\textbf
  {\bibinfo {volume} {107}},\ \bibinfo {pages} {060404} (\bibinfo {year}
  {2011})}\BibitemShut {NoStop}%
\bibitem [{\citenamefont {Gardiner}\ and\ \citenamefont
  {Zoller}(2004)}]{Gardiner2004}%
  \BibitemOpen
  \bibfield  {author} {\bibinfo {author} {\bibfnamefont {C.}~\bibnamefont
  {Gardiner}}\ and\ \bibinfo {author} {\bibfnamefont {P.}~\bibnamefont
  {Zoller}},\ }\href
  {http://www.springer.com/physics/quantum+physics/book/978-3-642-06094-6}
  {\emph {\bibinfo {title} {{Quantum Noise}}}},\ \bibinfo {edition} {4th}\ ed.\
  (\bibinfo  {publisher} {Springer},\ \bibinfo {year} {2004})\ p.\ \bibinfo
  {pages} {449}\BibitemShut {NoStop}%
\bibitem [{\citenamefont {Carter}\ \emph {et~al.}(1987)\citenamefont {Carter}
  \emph {et~al.}}]{Carter1987}%
  \BibitemOpen
  \bibfield  {author} {\bibinfo {author} {\bibfnamefont {S.~J.}\ \bibnamefont
  {Carter}} \emph {et~al.},\ }\href {\doibase 10.1103/PhysRevLett.58.1841}
  {\bibfield  {journal} {\bibinfo  {journal} {Phys. Rev. Lett.}\ }\textbf
  {\bibinfo {volume} {58}},\ \bibinfo {pages} {1841} (\bibinfo {year}
  {1987})}\BibitemShut {NoStop}%
\bibitem [{\citenamefont {Corney}\ \emph {et~al.}(2008)\citenamefont {Corney}
  \emph {et~al.}}]{Corney2008}%
  \BibitemOpen
  \bibfield  {author} {\bibinfo {author} {\bibfnamefont {J.}~\bibnamefont
  {Corney}} \emph {et~al.},\ }\href {\doibase 10.1103/PhysRevA.78.023831}
  {\bibfield  {journal} {\bibinfo  {journal} {Phys. Rev. A}\ }\textbf {\bibinfo
  {volume} {78}},\ \bibinfo {pages} {023831} (\bibinfo {year}
  {2008})}\BibitemShut {NoStop}%
\bibitem [{\citenamefont {Hoffmann}\ \emph {et~al.}(2008)\citenamefont
  {Hoffmann}, \citenamefont {Corney},\ and\ \citenamefont
  {Drummond}}]{Hoffmann2008}%
  \BibitemOpen
  \bibfield  {author} {\bibinfo {author} {\bibfnamefont {S.}~\bibnamefont
  {Hoffmann}}, \bibinfo {author} {\bibfnamefont {J.}~\bibnamefont {Corney}}, \
  and\ \bibinfo {author} {\bibfnamefont {P.~D.}\ \bibnamefont {Drummond}},\
  }\href {\doibase 10.1103/PhysRevA.78.013622} {\bibfield  {journal} {\bibinfo
  {journal} {Phys. Rev. A}\ }\textbf {\bibinfo {volume} {78}},\ \bibinfo
  {pages} {013622} (\bibinfo {year} {2008})}\BibitemShut {NoStop}%
\bibitem [{\citenamefont {Norrie}\ \emph {et~al.}(2006)\citenamefont {Norrie},
  \citenamefont {Ballagh},\ and\ \citenamefont {Gardiner}}]{Norrie2006}%
  \BibitemOpen
  \bibfield  {author} {\bibinfo {author} {\bibfnamefont {A.}~\bibnamefont
  {Norrie}}, \bibinfo {author} {\bibfnamefont {R.}~\bibnamefont {Ballagh}}, \
  and\ \bibinfo {author} {\bibfnamefont {C.}~\bibnamefont {Gardiner}},\ }\href
  {\doibase 10.1103/PhysRevA.73.043617} {\bibfield  {journal} {\bibinfo
  {journal} {Phys. Rev. A}\ }\textbf {\bibinfo {volume} {73}},\ \bibinfo
  {pages} {043617} (\bibinfo {year} {2006})}\BibitemShut {NoStop}%
\bibitem [{\citenamefont {Jack}(2002)}]{Jack2002}%
  \BibitemOpen
  \bibfield  {author} {\bibinfo {author} {\bibfnamefont {M.}~\bibnamefont
  {Jack}},\ }\href {\doibase 10.1103/PhysRevLett.89.140402} {\bibfield
  {journal} {\bibinfo  {journal} {Phys. Rev. Lett.}\ }\textbf {\bibinfo
  {volume} {89}},\ \bibinfo {pages} {140402} (\bibinfo {year}
  {2002})}\BibitemShut {NoStop}%
\bibitem [{\citenamefont {Egorov}\ \emph {et~al.}(2011)\citenamefont {Egorov}
  \emph {et~al.}}]{Egorov2011}%
  \BibitemOpen
  \bibfield  {author} {\bibinfo {author} {\bibfnamefont {M.}~\bibnamefont
  {Egorov}} \emph {et~al.},\ }\href {\doibase 10.1103/PhysRevA.84.021605}
  {\bibfield  {journal} {\bibinfo  {journal} {Phys. Rev. A}\ }\textbf {\bibinfo
  {volume} {84}},\ \bibinfo {pages} {021605(R)} (\bibinfo {year}
  {2011})}\BibitemShut {NoStop}%
\bibitem [{\citenamefont {Mertes}\ \emph {et~al.}(2007)\citenamefont {Mertes}
  \emph {et~al.}}]{Mertes2007}%
  \BibitemOpen
  \bibfield  {author} {\bibinfo {author} {\bibfnamefont {K.}~\bibnamefont
  {Mertes}} \emph {et~al.},\ }\href {\doibase 10.1103/PhysRevLett.99.190402}
  {\bibfield  {journal} {\bibinfo  {journal} {Phys. Rev. Lett.}\ }\textbf
  {\bibinfo {volume} {99}},\ \bibinfo {pages} {190402} (\bibinfo {year}
  {2007})}\BibitemShut {NoStop}%
\bibitem [{\citenamefont {Hall}\ \emph {et~al.}(1998)\citenamefont {Hall} \emph
  {et~al.}}]{Hall1998}%
  \BibitemOpen
  \bibfield  {author} {\bibinfo {author} {\bibfnamefont {D.}~\bibnamefont
  {Hall}} \emph {et~al.},\ }\href {\doibase 10.1103/PhysRevLett.81.4532}
  {\bibfield  {journal} {\bibinfo  {journal} {Phys. Rev. Lett.}\ }\textbf
  {\bibinfo {volume} {81}},\ \bibinfo {pages} {4532} (\bibinfo {year}
  {1998})}\BibitemShut {NoStop}%
\bibitem [{\citenamefont {Kaufman}\ \emph {et~al.}(2009)\citenamefont {Kaufman}
  \emph {et~al.}}]{Kaufman2009}%
  \BibitemOpen
  \bibfield  {author} {\bibinfo {author} {\bibfnamefont {A.}~\bibnamefont
  {Kaufman}} \emph {et~al.},\ }\href {\doibase 10.1103/PhysRevA.80.050701}
  {\bibfield  {journal} {\bibinfo  {journal} {Phys. Rev. A}\ }\textbf {\bibinfo
  {volume} {80}},\ \bibinfo {pages} {050701} (\bibinfo {year}
  {2009})}\BibitemShut {NoStop}%
\bibitem [{\citenamefont {Wineland}\ \emph {et~al.}(1994)\citenamefont
  {Wineland} \emph {et~al.}}]{Wineland1994}%
  \BibitemOpen
  \bibfield  {author} {\bibinfo {author} {\bibfnamefont {D.}~\bibnamefont
  {Wineland}} \emph {et~al.},\ }\href {\doibase 10.1103/PhysRevA.50.67}
  {\bibfield  {journal} {\bibinfo  {journal} {Phys. Rev. A}\ }\textbf {\bibinfo
  {volume} {50}},\ \bibinfo {pages} {67} (\bibinfo {year} {1994})}\BibitemShut
  {NoStop}%
\bibitem [{\citenamefont {S{\o}rensen}\ \emph {et~al.}(2001)\citenamefont
  {S{\o}rensen} \emph {et~al.}}]{Sorensen2001}%
  \BibitemOpen
  \bibfield  {author} {\bibinfo {author} {\bibfnamefont {A.}~\bibnamefont
  {S{\o}rensen}} \emph {et~al.},\ }\href {\doibase 10.1038/35051038} {\bibfield
   {journal} {\bibinfo  {journal} {Nature}\ }\textbf {\bibinfo {volume}
  {409}},\ \bibinfo {pages} {63} (\bibinfo {year} {2001})}\BibitemShut
  {NoStop}%
\bibitem [{\citenamefont {Riedel}\ \emph {et~al.}(2010)\citenamefont {Riedel}
  \emph {et~al.}}]{Riedel2010}%
  \BibitemOpen
  \bibfield  {author} {\bibinfo {author} {\bibfnamefont {M.~F.}\ \bibnamefont
  {Riedel}} \emph {et~al.},\ }\href {\doibase 10.1038/nature08988} {\bibfield
  {journal} {\bibinfo  {journal} {Nature}\ }\textbf {\bibinfo {volume} {464}},\
  \bibinfo {pages} {1170} (\bibinfo {year} {2010})}\BibitemShut {NoStop}%
\bibitem [{\citenamefont {Gross}\ \emph {et~al.}(2010)\citenamefont {Gross}
  \emph {et~al.}}]{Gross2010}%
  \BibitemOpen
  \bibfield  {author} {\bibinfo {author} {\bibfnamefont {C.}~\bibnamefont
  {Gross}} \emph {et~al.},\ }\href {\doibase 10.1038/nature08919} {\bibfield
  {journal} {\bibinfo  {journal} {Nature}\ }\textbf {\bibinfo {volume} {464}},\
  \bibinfo {pages} {1165} (\bibinfo {year} {2010})}\BibitemShut {NoStop}%
\bibitem [{\citenamefont {Drummond}\ \emph {et~al.}(1993)\citenamefont
  {Drummond} \emph {et~al.}}]{Drummond1993a}%
  \BibitemOpen
  \bibfield  {author} {\bibinfo {author} {\bibfnamefont {P.~D.}\ \bibnamefont
  {Drummond}} \emph {et~al.},\ }\href {\doibase 10.1038/365307a0} {\bibfield
  {journal} {\bibinfo  {journal} {Nature}\ }\textbf {\bibinfo {volume} {365}},\
  \bibinfo {pages} {307} (\bibinfo {year} {1993})}\BibitemShut {NoStop}%
\end{thebibliography}%

\end{document}